\documentclass[acus]{JAC2003}

\usepackage{graphicx}
\usepackage{booktabs}

\begin{document}
\title{A NEW APPROACH TO LINAC RESONANCES AND  EQUIPARTITION ? \\[-.7\baselineskip]}

\author{I. Hofmann, GSI-HI-Jena, 64291 Darmstadt, Germany}
\maketitle

\begin{abstract}
In this note we refer to a recent paper ``Equipartition, Reality
or Swindle?''~\cite{lagniel2012} presented by Lagniel at HB2012,
Beijing,  which claims to challenge the currently used approach to
describe space charge resonances and emittance exchange with the
help of linac-specific stability charts. On the one hand we agree
with the general observation that enforcing equipartition (EP)
would be an unnecessary constraint; however, we find that the
heuristic single-particle arguments and examples presented by
Lagniel are speculative and cannot be reconciled with results from
self-consistent computer simulation. Thus,  we see no
justification for Lagniel's suggestions, which include a modified
EP definition (treating $x$ and $y$ as correlated). Instead, we
suggest to maintain the current approach and to continue using the
``conventional'' EP definition. With our findings we also respond
in some detail to Lagniel's ``Topics of
 Discussion''.
\end{abstract}


\section{Introduction}
One of the widely accepted criteria in high-current linac design
is to use linac-specific stability charts to identify parameter
regions, where emittance exchange between the longitudinal and
transverse degrees of freedom might occur. It is common
understanding that this exchange is caused by space charge
resonances. Plotting the rms tune ratios and tune depressions from
simulation codes on these stability charts is a widely used
approach in new linac projects in order to deal with the problem
of undesirable emittance exchange. These charts include the
physics of phase space flow on a perturbational level, and under
the effect of coupling between two degrees of freedom due to space
charge ``pseudo-multipoles''~\cite{hofmann1998}.  The resulting
resonance stop-bands proliferate not only the location of possible
resonances lines (as ring resonance diagrams normally do), they
also represent their space charge dependent driving terms.
Obviously, particle-in-cell (PIC) simulation  is necessary to
examine the validity of the charts, which was carried out
successfully under a great variety of conditions (for a recent
discussion including relevant references see
Ref.~\cite{hofmann-hb2012b}).

Lagniel's paper is based on three arguments mainly:

(1)  The law of EP in a rigorous sense holds only for ergodic
systems (undeniable - see our comments in the before last
section).

(2) If used at all, the ``conventional'' EP condition employing an
rms energy ratio $T$ as shown in Eq.~\ref{anis}
\begin{equation}
T\equiv \frac{\epsilon_z k_z}{\epsilon_x k_x}=1, \label{anis}
\end{equation}
(all quantities understood as rms quantities) \textit{``is
wrong''} and should have a factor 2 in the denominator, with the
argument that the {\it{sum}} of transverse energies should be in
balance with the longitudinal energy (see Eq. (5) in
Ref.~\cite{lagniel2012}).

(3) The conventional EP-condition does not prevent resonant
emittance exchange.

Below we examine first assertion (3) in the next section, followed
by a critical discussion of (2), and concluding with comments on
Lagniel's ``Topics of discussion'' as well as some final remarks.

\section{PIC examination of Lagniel's examples}\label{examples}

In the following we undertake a careful examination of the two
examples of equipartitioned and non-equipartitioned beams by using
the TRACEWIN particle-in-cell simulation of realistic bunched
beams in a periodic   FODO lattice, with no acceleration and RF
gaps to keep the bunches longitudinally. We employ 100.000
simulation particles and an input distribution following the
TRACEWIN standard option of randomly generated particles in the 4d
transverse hyper space as well as randomly in the longitudinal
phase plane ellipse.

\begin{figure}[h]
   \centering
   \includegraphics*[width=75mm]{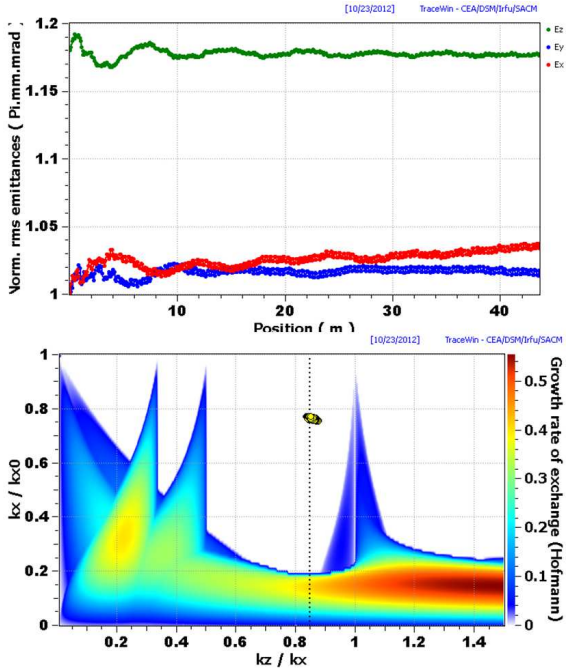}
   \caption{Evolution of rms emittances for example on the EP condition, with stability chart for $\epsilon_z/\epsilon_x =1.18$ (dotted line is EP condition $T=1$).}
   \label{both1}
\end{figure}

The parameters of the exactly equipartitioned example ($T=1$ by
the conventional definition of Eq.~\ref{anis}) described in Eqs.
(6)-(9) of Ref.~\cite{lagniel2012} are identically chosen as
$\epsilon_z/\epsilon_x =1.18$,  $k_z/k_x =0.85$ ,  $k_ x/k_{0x}
=0.75$ and $k_{0z} =73^o$ (per cell). The beam is matched - as
much as possible with TRACEWIN - and then propagated over 100
cells (45 m). The result  is shown in Fig.~\ref{both1} including
the tune footprint from TRACEWIN on the stability chart for
$\epsilon_z/\epsilon_x =1.18$. Note that the vertical line at
$k_z/k_x =1$ (the ``main'' or ``fourth order'' resonance
$2k_{z}-2k_{x}=0$) in the stability charts corresponds to the
diagonal in Lagniel's Figure1.
It is noted that there is a 2-3\% initial transverse emittance growth,
whereas the longitudinal emittance is  only oscillating around its
initial value (1.18).  Running this case over twice the distance  we  find that $\epsilon_z$ still remains constant within $\pm 0.1\%$ and  $\epsilon_x, \epsilon_y$ grow slowly, but $<1\%$.  Hence, we find absence of emittance exchange - consistent with the ``safe'' distance of the tune footprint from the stop-band in the stability chart. Note here that the
width of the stop-band of the main resonance (near $k_z/k_x =1$) is
shrinking to zero for $\epsilon_z/\epsilon_x \rightarrow
1$ (see Ref.~\cite{hofmann-hb2012b}).

Thus, contrary to Ref.~\cite{lagniel2012}, we find that the proximity of the
EP working point to the main resonance $k_{z}/k_{x}=1$  is not adversary to the
stability of the rms emittances. In fact, Lagniel is drawing his conclusions
from  a schematic picture of a tune footprint in his Figure 1,
which he finds indicative of an overlap with the resonance $k_{z}/k_{x}=1$. Firstly,
the square box tune footprint is un-physical - particles not
seeing any space charge in one direction (as the ones on
the two sides adjacent to the right upper corner, which is given by the
zero-current tunes) don't exist. Consistent tune footprints in a $k_z - k_x$ plane are
actually necktie-shaped rather than square-boxed, which reduces the overlap. Secondly,
it is necessary to consider the response of the bunch as a whole, i.e. the property of the ensemble versus that of single particles. Individual particles may have
growing amplitudes in one direction, which can be compensated by other particles with shrinking amplitudes, and no net effect.

\begin{figure}[h]
   \centering
   \includegraphics*[width=75mm]{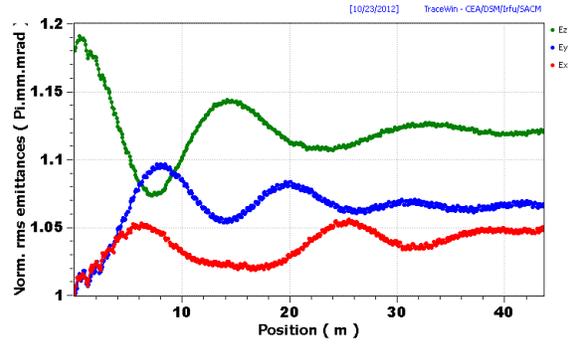}
   \caption{Same as Fig.~\ref{both1}, but $k_{z}/k_{x}=1.02$.}
   \label{emit5}
\end{figure}

Not surprisingly there is a fast ($<20$ cells) and pronounced
emittance exchange of about 10\% , if we
 lower  the transverse focusing such that $k_{z}/k_{x}=1.02$ and the working point sits exactly on the stop-band,
   while the beam is initially weakly non-equipartitioned with $T=1.2$
   (Fig.~\ref{emit5}).
As far as the second example, following Eq. (10)  of Ref.~\cite{lagniel2012} (with  $k_{0x}$ further lowered to 50$^o$), we agree, in principle, that no emittance exchange should be expected. The distance to the fourth order resonance line $k_{z}/k_{x}=1$  is large enough, equally to the third order  line $k_{z}/k_{x}=2$ (i.e. $k_{z}-2k_{x}=0$).   It should be noted, however, that the driving term for this third order mode requires a sextupolar component in the space charge potential. Such a third order term is absent in the matched initial beam, where only even powers in the space charge potential exist. Therefore we cannot see how it enters into Lagniel's frame of discussion, if the driving term is absent. In PIC simulation however, equally in our stability chart, this driving term evolves self-consistently from a {\it resonant unstable behaviour} of the third order mode building up from initial noise. An example for the effect of this third order mode on the extended stability chart  including $k_{z}/k_{x}=2$ is given below in Fig.~\ref{both16}.

\section{Do we need a new EP-formula?}\label{EP}

In the discussion preceding his Eq.~(5), Lagniel argues that the conventional EP condition Eq.~\ref{anis} ``is wrong'' and should have a factor 2 in the denominator to account for ``a total correlation between the two radial degrees of freedom''.
We cannot follow this argument,  because dynamically speaking {\it each} degree of freedom is independent - no matter what its initial tune and emittance values are. But let us use simulation to help decide between the conventional EP condition and Lagniel's proposition. To this end let us call Lagniel's modified energy ratio $T^*$ ($=T/2$ and the condition $T^*=1$ the modified equipartition condition EP$^*$.

\begin{figure}[h]
   \centering
   \includegraphics*[width=75mm]{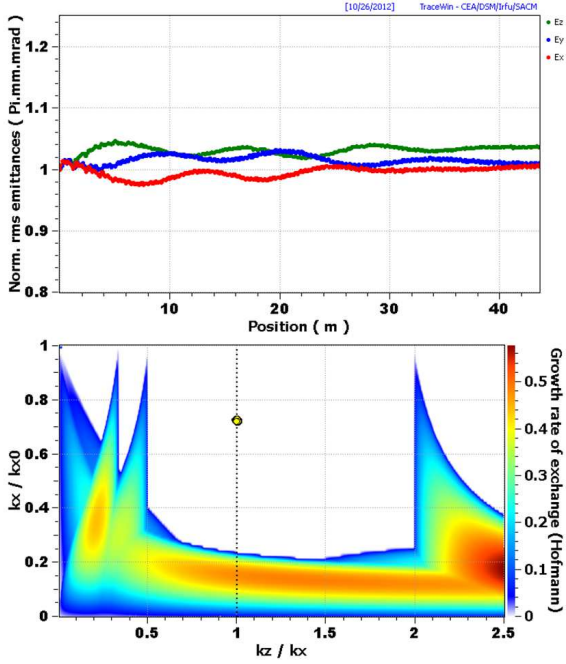}
   \caption{Case 1: Evolution of rms emittances and stability chart for   $\epsilon_z/\epsilon_x =1$ (dotted line is EP condition $T=1$).}
   \label{both13}
\end{figure}

Let us start with Case 1: fulfilling the ``conventional'' EP condition  ($T=1$, but  $T^*=1/2$), with
 $\epsilon_z/\epsilon_x =1$,  $k_z/k_x =1$ (also  $k_{0z}/k_{0x} =1$)
and $k_{0z} =73^o$ per cell as well as a transverse tune depression of  $k_{x}/k_{0x} =0.72$.  The resulting  rms emittances as well as the footprint of tunes on the stability chart are shown in
Fig.~\ref{both13}.
Besides the usual fluctuations there is no real emittance transfer - consistent with the stability chart.  Note that there is an indication that the two transverse emittances actually behave as independent and undergo small deviations varying in time - in spite of ``identical'' starting conditions.

\begin{figure}[h]
   \centering
   \includegraphics*[width=75mm]{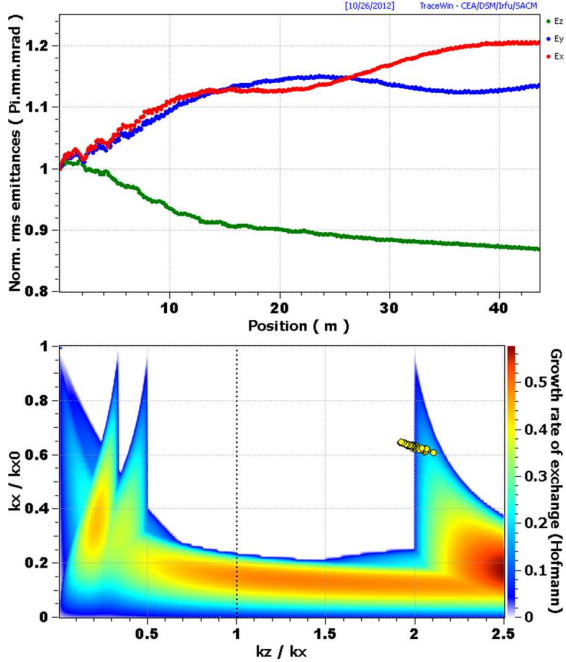}
   \caption{Case 2: Weaker transverse focusing and stability chart for   $\epsilon_z/\epsilon_x =1$.}
   \label{both16}
\end{figure}

Now we switch to Case 2: a weaker transverse focusing (but same emittance ratio), such that $k_z/k_x =2$ and EP$^*$ is fulfilled ($T^*=1$, whereas $T=2$ ).   Results are shown in Fig.~\ref{both16}.
There is  a real emittance transfer from the longitudinal direction into transverse - obviously induced by the third order resonance discussed in the end of Section~\ref{examples}.  For the (conventional) energy ratio we find $T=2\to  1.3$. Note that the ``independent'' behaviour of transverse emittances is even more pronounced than in Fig.~\ref{both13}, hence Lagniel's argument of ``total correlation'' between $x$ and $y$ is not supported.

We have also examined a Case 3: with $\epsilon_z/\epsilon_x =2$
and  $k_z/k_x =1$ we find the expected main resonance, which leads
to a fast emittance transfer during the first 10 cells already,
with a final evolution  $T=2 \rightarrow  1.2$. Below we summarize
the impact of these three simulations on the
energy ratios $T$ as well as $T^*$:\\

$
\begin{array}{cccc}
              & Case  1                   & Case  2                    & Case 3\\
  T          &  1\rightarrow 1          &  2 \rightarrow 1.3      &  2 \rightarrow  1.2\\
  T^*  &  0.5 \rightarrow 0.5            &  1 \rightarrow 0.65     &  1 \rightarrow 0.6\\
\end{array}
$
\\

In view of all this we find it straightforward to continue with
the conventional definition of EP as $T=1$ (using $T$ defined in
Eq.~\ref{anis}), which considers all degrees of freedom as
independent. This is supported by the ``splitting'' of transverse
emittances; furthermore by the fact that we have not found (by
simulation, and avoiding extreme tune depression) a single case,
where $T=1$ is subject to emittance transfer - in contrast with
the assertions in Ref.~\cite{lagniel2012}. The initial $T^*=1$ as
in Case 2, instead, is unstable.

\section{Lagniel's Topics of Discussion}
Based on the above findings we attempt to respond to the
discussion opened in the last section of Lagniel's paper by
referring to his six points and starting with the \textit{original
quotations} from Ref.\cite{lagniel2012}.

1- \textit{"The linac beams are out of the EQP theorem validity
limit, to apply the "EQP rule" designing a linac is a mistake."}
It is undeniable that ``true'' equipartition can be applied  to
ergodic systems only. As we have a general difficulty to describe
and measure distributions in 6D phase space, the concept of
projections into 2D planes and of rms values in 2D was developed -
successfully so far. In the same spirit it has become common
practice to employ an \textit{rms energy ratio} ($T$ in
Eq.\ref{anis} as a reduced, but well-defined quantity) and call
the special case $T=1$ equipartitioned. Whether or not $T=1$ is a
practically helpful requirement is a different question.

2- \textit{"The application of the "EQP rule" does not prevent
emittance exchanges induced by coupling resonances."} We find this
statement is a speculative interpretation of fictitious beam
footprints and \textit{not} supported by our PIC simulation, also
not by the stability charts. We have simulated exactly the same
case as in Lagniel's example and find that definitely
  no rms emittance exchange occurs (similarly for a variety of other
  initial emittance ratios, still equipartitioned).

3- \textit{"Safe tunes with beam footprints out of the coupling
resonances can be found when the "EQP rule" is not respected."} -
a well-established recognition in the linac community.

4- \textit{"The constraint imposed by the "EQP rule" on a linac
design can lead to a non optimized beam dynamics and higher
construction and operation costs."} - out of question.

5- \textit{"The question of energy exchange / emittance transfer
must be analyzed as done in circular machines (tune diagram,
evaluation of the resonance' excitation strength)."} Authors
should feel free to introduce different kinds of tune diagrams as
long as they  prove  they are viable. We have suggested linac
stability charts as they include tune depression  (intensity)
\textit{and}
 tune ratios. Circular machine diagrams with $k_z$,  $k_x$
(or $k_y$) separate may be fine, but would require a third
dimension to include intensity. Actually, as we need to worry only
about difference resonances of the kind $nk_z-mk_x =0$, the tune
ratios suffice. The resonance driving terms are already part of
the stop-band widths of the stability charts and need not to be
evaluated separately.

6- \textit{"The modern physics tools developed to characterize the
level of disorder (chaos) present in nonlinear Hamiltonian systems
could be applied to characterize and optimize our beams."} It
should  certainly be welcomed to continue using all the great
tools developed in nonlinear dynamics.

Finally, we would like to comment also on Lagniel's question at
the end of his before last section: \textit{"Why the belief in EQP
did not pollute the synchrotron world?"} Synchrotrons indeed have
many resonances to worry  about. However, as early as 1968,
Montague already warned about the effect of horizontal-vertical
emittance exchange by a space charge pseudo-octupole resonance on
the main diagonal of the CERN Proton Synchrotron
($2Q_x-2Q_y=0$)~\cite{montague}. Owed to its possible importance
for high-current operation at CERN, the subject was carefully
studied experimentally only much later - with excellent agreement
with theory~\cite{metral}.

\section{Final Remarks}
We have shown that Lagniel's assertions on EP and on linac
resonances are not supported by  PIC simulations, therefore a new
approach to this topic on the ground of the presented arguments
cannot be seen. Independent of this it is known since many years
that there is no necessity to enforce EP, as most of the parameter
space is filled by non-equipartitioned regions, where no emittance
coupling is found - as shown by the linac stability charts.

It should be emphasized here that the notion of EP or non-EP in
our context is based on rms quantities (emittances, tunes). Such
an approach obviously cannot say anything about the question -
also raised by Lagniel - of energy equipartition on surfaces in a
multi-dimensional phase space.  It would certainly be welcomed by
everybody if future analysis would go beyond rms measures, for
example including halo distributions and the question of coupling
in the tail distributions, and thus open a new dimension of this
problem to the scientific discussion. At the time being, however,
linac designers may continue to work with their validated tools
and feel free to be on EP, or not to be on EP - as long as they
have a convincing reason for it.
\\
\\
{\bf Acknowledgment:} The author is indebted to G. Franchetti  for
valuable discussions.


\begin{thebibliography}{99}
\bibitem{lagniel2012}  J.-M. Lagniel, HB2012 conference, Beijing, paper TUO3A03 (2012)
\bibitem{hofmann1998}I. Hofmann, {\em Phys. Rev.} {\bf E 57}, 4713 (1998).
\bibitem{hofmann-hb2012b} I. Hofmann, HB2012 conference, Beijing, paper TUO3A01 (2012)
\bibitem{montague}B.W. Montague, CERN-Report No. 68-38, CERN (1968)
\bibitem{metral}E. Metral et al., Proc. of EPAC 2004, p. 1894 (2004)
\end{thebibliography}
\end{document}